\documentstyle[12pt]{article}
\newcommand{\nc}{\newcommand}
\nc{\beq}{\begin{equation}} \nc{\eeq}{\end{equation}}
\nc{\beqa}{\begin{eqnarray}} \nc{\eeqa}{\end{eqnarray}}
\nc{\lsim}{\begin{array}{c}\,\sim\vspace{-21pt}\\< \end{array}}
\nc{\gsim}{\begin{array}{c}\sim\vspace{-21pt}\\> \end{array}}

\nc{\al}{\alpha}
\nc{\Q}[2]{Q_{#1#2}}
\nc{\Vb}{{\bar V}}
\nc{\q}[2]{q_{#1#2}}

\nc{\N}{{\bar N}}
\nc{\qb}{{\bar q}}
\nc{\hb}{{\bar h}}
\nc{\Nf}{{N_f}}
\nc{\su}{SU(\N)}
\nc{\F}{F_h} 

\nc{\alphae}{\alpha_e}
\nc{\am}{\alpha_m}
\nc{\ay}{\alpha_{\lambda}}
\nc{\Mg}{\tilde{M}_{\tilde{g}}}
\nc{\msq}{\tilde{m}_q^2}
\nc{\msqb}{\tilde{m}_{\bar{q}}^2}
\nc{\mm}{\tilde{m}_M^2}
\nc{\ao}{\alpha_0}
\nc{\af}{\alpha_*}
\nc{\g}{g_{\lambda}}
\nc{\f}{f_{\lambda}}
\begin{document}

\begin{titlepage}
\begin{center}
\vspace{2cm}
{\hbox to\hsize{\hfill Fermilab-Pub-97/420-T }}
{\hbox to\hsize{\hfill WIS-98/1/Jan-PP }}

\bigskip

\vspace{2cm}
\bigskip
\bigskip
\bigskip
{\Large \bf   Duality in the Presence of Supersymmetry Breaking
}

\bigskip
\bigskip
{\bf Hsin-Chia Cheng}$^{\bf a}$ and {\bf Yael Shadmi}$^{\bf b}$
\\

\bigskip

\bigskip

$^{\bf a}${ \small \it Fermi National Accelerator Laboratory\\
  P.O.Box 500, Batavia\\
  IL 60510, USA\\

 \smallskip

{\rm email}: hcheng@fnal.gov\\ }

\bigskip

$^{\bf b}${ \small \it Department of Particle Physics\\
  Weizmann Institute of Science\\
  Rehovot 76100, Israel\\

 \smallskip

{\rm email}: yshadmi@wicc.weizmann.ac.il \\
}
\vspace{1.3cm}
{\bf Abstract}

\end{center}

\noindent 
We study Seiberg duality for ${\cal N}=1$ supersymmetric QCD with
soft supersymmetry-breaking terms. We generate the soft terms 
through gauge mediation by coupling two theories related by Seiberg 
duality to the same supersymmetry-breaking sector. In this way, 
we know what a supersymmetry-breaking perturbation in one theory 
maps into in its ``dual''.  Assuming a canonical K\"ahler potential 
we calculate the soft terms induced in the magnetic theory and
find that some of the scalars acquire negative masses squared.
If duality is still good for small supersymmetry breaking, 
this may imply some specific symmetry breaking patterns for 
supersymmetric QCD with small soft supersymmetry-breaking masses,
in the case that its dual theory is weakly coupled in the infrared.
In the limit of large supersymmetry breaking, the electric
theory becomes ordinary QCD. However, the resulting symmetry breaking 
in the magnetic theory is incompatible with that expected for QCD.

\end{titlepage}

\renewcommand{\thepage}{\arabic{page}}
\setcounter{page}{1}

\baselineskip=18pt

\section{Introduction }  %
\label{introduction}

The constrained structure of supersymmetric (SUSY)
field theories provides powerful tools for analyzing
 their strong dynamics.
Using these tools Seiberg gave striking evidence for
the existence of ``dual" pairs of ${\cal N}=1$ supersymmetric gauge
 theories that give the same infrared (IR) physics~\cite{Seiberg}.
One is immediately led to ask whether a similar phenomenon
exists in the absence of 
supersymmetry~\cite{Hsu}-\cite{DHoker},
and in particular, whether theories related by Seiberg duality in the
supersymmetric limit still give the same infrared physics when 
supersymmetry is 
broken~\cite{terninganomaly}. 
To answer this question, one obviously starts with a pair
of dual ${\cal N}=1$ supersymmetric 
theories\footnote{Refs.~\cite{A-G} consider a softly broken 
 ${\cal N}=2$ $SU(2)$ theory and use the symmetries of the Seiberg-Witten
solution~\cite{seibergwitten} to study the low-energy theory,
but only for $N_f \le 2$.}.
The question of how to introduce supersymmetry breaking into
this system is less obvious. 
In~\cite{Peskin}, soft supersymmetry breaking masses were added
 to both theories with the assumption that the scalar masses squared
are positive. This assumption was relaxed in~\cite{DHoker}.
In Refs.~\cite{Hsu,Hsu2},
soft masses were obtained by promoting some couplings to
spurion fields, with frozen supersymmetry-breaking vacuum expectation
values (vevs). 
It is not clear however what an explicit, supersymmetry-breaking
perturbation in one theory maps into in the dual theory.
In fact, even in the supersymmetric case, only the chiral operator map
between the two theories is known in 
general~\cite{Seiberg,nonchiralmap}.
Alternatively, one may study theories in which supersymmetry is only
spontaneously broken\footnote{A toy model with spontaneous 
supersymmetry breaking was used in the first paper of~\cite{Hsu} to justify the 
use of spurion fields to generate soft terms.}. 
Then the Lagrangian  is manifestly
supersymmetric, and one has some confidence in mapping superpotential
perturbations between the two theories. 
This is the approach we take in this paper.

We start with an $SU(N)$ gauge theory with $F$ flavors of matter
in the fundamental representation,
and imagine coupling it to a sector with dynamical supersymmetry breaking (DSB).
Such theories have been extensively studied in the context of
gauge-mediated supersymmetry breaking~\cite{GMSB,dnns,review}.
Some $SU(N)$ supermultiplets couple directly to the DSB sector
and become heavy with supersymmetry-breaking mass splittings.
Soft masses are then induced for the remaining $SU(N)$ squarks
and gluinos through loops involving the heavy fields. 
In the supersymmetric limit, which is typically attained by setting
some superpotential couplings in the DSB sector to zero,
the theory has a dual description with gauge group $SU(F-N)$
and with the DSB sector essentially unchanged.
We then turn supersymmetry breaking back on.
This amounts to adding some superpotential term involving
the DSB fields in the electric theory.
(We will loosely refer to the original theory as
the electric theory, and to the theory obtained by the duality operation
as the magnetic theory.)
Since supersymmetry is only dynamically broken, 
we know what this term maps into in the magnetic theory.
We then study the effect of the perturbation on the magnetic theory.
In this theory, some fields are coupled directly to the DSB sector
and become heavy with supersymmetry-breaking mass splittings.
As in the electric theory, soft terms are then induced for the
 remaining $SU(F-N)$ fields, through loops involving the heavy fields.
In some cases, and under certain assumptions, we will be able to calculate 
these masses. 
We will mainly be interested in the sign of the mass-squared
induced for the scalar fields of the magnetic theory.
As we will see, these are often negative, resulting in definite 
predictions for the pattern of global symmetry breaking
 in the low energy theory.

For small soft masses our analysis is reliable only when the magnetic theory 
is weakly coupled at low energies.
Then the low-energy electric theory is strongly coupled, 
and we have no direct information about its behavior in the presence of small
supersymmetry breaking.
If duality continues to hold in the presence of small breaking,
our findings for the vacuum of the magnetic theory then give
a prediction for the pattern of chiral symmetry breaking in the electric
 theory.

For large supersymmetry breaking, we compare the global symmetry of the
magnetic theory at low energies to what we expect in the electric theory.
In many cases the two are incompatible, indicating that the
proposed duality (with the assumptions we have made) breaks down.

The duality transformation was shown to withstand different perturbations.
The electric and magnetic theories flow to the same 
IR theory after a mass term is added~\cite{Seiberg}, 
or when a global symmetry is gauged~\cite{PST}.
As long as supersymmetry is preserved, one does not expect any 
phase transitions to occur as the
size of the perturbation is 
increased~\cite{seibergwitten,nophasetransition}.
Once supersymmetry is broken, this is no longer the case.
The electric and magnetic theories may undergo a phase transition and
cease to be equivalent.

A crucial ingredient in our analysis is the
K\"ahler potential of the dual theory,
which is not known even in the supersymmetric case.
In that case, the theories are only known to agree at zero energy, 
and the details of the K\"ahler potential
are irrelevant for this agreement to hold.
Here we study a simple class of candidate duals, those with a canonical
K\"ahler potential, up to field renormalizations. 
These theories may not be the ``real'' duals. 
In fact, the 
correct duals could be ones with complicated  K\"ahler potentials.
Our analysis and results thus only apply to this class of simple candidate
duals.
This problem also plagues previous attempts to study
non-supersymmetric duality.
However, with the assumption about the K\"ahler potential in place,
we can actually calculate the soft terms in both the electric 
and the magnetic theory,
instead of adding squark masses by hand.

This paper is organized as follows.
In section~\ref{framework} we present the model we study and discuss
our assumptions. 
We then calculate the soft masses induced in the magnetic theory
for small supersymmetry breaking in section~\ref{smallbreaking}.
We consider separately the case that the magnetic theory is  
infrared-free, and the case where it is at a Banks-Zaks fixed point.
Finally, we calculate the baryon and meson 
soft masses in the case $N_f=N_c+1$ through 
a completely higgsed dual description. 
In section~\ref{vacuum} we study the implications of the soft masses
we found for the pattern of chiral symmetry breaking
in the vacuum of the magnetic theory.
We then move on to the case of large supersymmetry breaking 
in section~\ref{largebreaking}.
Again, we present the soft masses generated in the
magnetic theory and the resulting global symmetry.
Section~\ref{conclusions} contains our conclusions and some final remarks.
In appendix~A we briefly discuss
the magnetic theory with pure matter messengers. 
Finally, some details of the calculation are summarized in appendix~B.

\section{Framework    }  %
\label{framework}

Let us now describe in more detail the model we consider.
The theory we start with is an $SU(N)$ gauge theory with $\Nf+1$ flavors:
the fields $Q^i$, $H$, transform as $SU(N)$ fundamentals, and the fields
${\bar Q}_i$, ${\bar H}$ transform as $SU(N)$ antifundamentals, with 
$i=1\ldots \Nf$.
The theory is coupled to a DSB sector 
through the superpotential coupling~\cite{dnns} 
\beq
\label{w}
S H\cdot {\bar H} \ ,
\eeq   
where $S$ is a field of the DSB sector. 
We do not specify here the details of the DSB sector.
Rather, in the following, we will clarify the different requirements 
on this sector. 
We return to this point at the end of this section.
We would like the full theory to have a stable minimum
in which none of the $SU(N)$ fields develop vevs at the tree level.
At this minimum, we would like the field
$S$ to develop $A$- and $F$- type vevs, which we denote by 
$S_0$ and $F_0$, with $S_0^2 > F_0$.
This can be ensured by taking the parameter that induces supersymmetry breaking to be small enough.
Then, the squarks and gluinos of $SU(N)$ acquire masses
of the order $F_0/S_0$ times a loop factor~\cite{dnns}.
Here we have implicitly assumed that the $SU(N)$ gauge theory is weakly coupled at the scale $S_0$.
Thus, if $SU(N)$ is asymptotically free, we take its scale $\Lambda$ to be much smaller than $S_0$.
If $SU(N)$ is not asymptotically free, we take $\Lambda$ to be the largest scale in the problem.

In the supersymmetric limit, the theory has a dual description with gauge group
$SU(\Nf+1-N)$,
fields 
$q_i$, $h$ and $\qb^i$, $\hb$ in the fundamental and antifundamental 
representations respectively, 
and gauge singlet fields 
$M^i_j$, $V^i$, $\Vb_i$ and $P$
corresponding to the mesons of the
electric theory~\cite{Seiberg}.
For the two theories to agree in the infrared, 
the following superpotential is required in the
magnetic theory~\cite{Seiberg}:
\beq
\label{wm}
W \ =\ {1\over \mu^\prime}\, 
\left( M^i_{j} q_i\cdot \qb^j \ +\
V^i q_i\cdot\hb \ +\
\Vb_i  h\cdot\qb^i \ +\
P h\cdot \hb \right) \ + \
P\, S \ +\ 
W_{SB}(S,\phi)
\ ,
\eeq
where the scale $\mu^\prime$ is required on dimensional grounds, and is related to the $SU(N)$
scale $\Lambda$ and the  $SU(\Nf+1-N)$ scale ${\bar \Lambda}$
through~\cite{Seiberg}
\beq
\label{scales}
\Lambda^{3N-\Nf-1}\, {\bar\Lambda}^{2(\Nf+1)-3N}\ \sim\ {\mu^\prime}^{\Nf+1}\ .
\eeq
The DSB sector remains untouched by the duality transformation.
In~(\ref{wm}) we indicated 
the superpotential associated with this sector, $W_{SB}(S,\phi)$,
and $\phi$ collectively denotes
the fields of this sector apart from $S$.

Redefining the fields to get a canonical K\"ahler potential, the superpotential
can be rewritten as,
\beq
\label{wm1}
W \ =\ \lambda\, 
\left( M^i_{j} q_i\cdot \qb^j \ +\
V^i q_i\cdot\hb \ +\
\Vb_i  h\cdot\qb^i \ +\
P h\cdot \hb \right) \ + \
\mu\, P\, S \ +\ 
W_{SB}(S,\phi)
\ ,
\eeq
where $\lambda$ is a dimensionless coupling.
The potential is then:
\beqa
\label{potential}
V \ &=&\ 
\lambda^2\,\Big( 
{\left\vert \, M^i_{j} \, q_i + \Vb_j\, \hb\,\right\vert}^2 \ + \ 
{\left\vert \, M^i_{j}\,  \qb^j + V^i\,  h\, \right\vert}^2 \ + \ 
{\left\vert\, q_i\cdot \qb^j \,\right\vert}^2  \ + \ 
{\left\vert\, q_i\cdot \hb\,\right\vert}^2  \ + \
{\left\vert\,  h\cdot \qb^j\,\right\vert}^2 \nonumber \\
 \ &+& \ 
{\left\vert \, \Vb_i\, \qb^i\ + \ P \hb \, \right\vert}^2  \ + \
{\left\vert\, V^i \,\q_i\ +\ P\, h \, \right\vert}^2  
\Big) \ + \
{\left\vert\, \lambda\, h\cdot \hb + \mu\, S \,\right\vert}^2 \nonumber
\\
 \ &+& \ 
{\left\vert \,\mu \,P + {\partial W_{SB}\over \partial S}\, \right\vert}^2 
+ V_{SB}(S,\phi) \ .
\eeqa
We would like to consider a minimum at which the fields $M^i_{j}$, $V^i$,
$\Vb_i$, $q_i$ and $\qb^i$ do not get tree-level vevs.
Extremizing the potential with respect to $P$ we thus find,
\beq
\label{peq}
\lambda^2 P\, ( {\vert h \vert}^2 + {\vert \hb \vert}^2 ) +
\mu\,\left(\mu\, P +  {\partial W_{SB}\over \partial S} \right)\,
\ =\ 0 \ .
\eeq
And for non-zero $\partial W_{SB}/ \partial S$, $P$ develops a non-zero vev.
The $h$ derivative then gives, 
\beq
\label{min}
\lambda\, {\vert P\vert}^2\, h^*\ + \ ( \lambda\, h\cdot\hb + \mu\, S )^*\, \hb
\ =\ 0 \ .
\eeq
Assuming that $S$ develops a vev, 
there are two qualitatively different possibilities then for a stable minimum. 
One is that $h$ 
and $\hb$ do not acquire vevs and the energy, apart from $ V_{SB}(S,\phi)$,
equals $\mu^2\, S^2$.
We will refer to this type of minimum as a ``matter messenger'' minimum.
Another possibility, which we will term a ``gauge messenger'' minimum, is that 
$h$ and $\hb$ develop vevs to compensate for the $S$ vev.
Clearly, when the $S$ vev is large compared to the amount of 
supersymmetry breaking,
this will be the preferred minimum.
This is also the minimum that connects smoothly to the supersymmetric case
when SUSY breaking vanishes.
If the dual description makes sense at all, 
it had better make sense for small breaking.
We will therefore focus on the gauge-messenger minimum.
The case of a magnetic theory with matter messengers will be 
discussed in appendix~A.

At the minimum we consider $h$ and $ \hb$ acquire vevs. We can choose
\beq
\label{hvevs}
h^{\N+1}\ = \ -\,\hb_{\N+1}\ =\ v \ ,
\eeq
with $\N  \equiv \Nf-N$.
The field $P$ also acquires a vev with
\beq
\label{veveqn}
\lambda\, P^2 \ +\ \lambda\, v^2\ -\ \mu \, S\ =\ 0\ .
\eeq
The $F$ components of $P$ and $h$ are then,
\beq
\label{fvevs}
F_h\ =\ \lambda P\, v\ ,\ \ \ F_P\ =\ - \lambda\, P^2\ .
\eeq

The $h$ vevs higgs the group down to $SU(\N)$. 
This is precisely what one would expect if there were no 
supersymmetry breaking in the theory.
A non-zero vev for $S$ in the electric theory is a mass term for one flavor
of that theory.
The dual theory is then higgsed by one unit.
At low energies, the electric theory is an $SU(N)$ theory with $\Nf$ 
flavors and the magnetic theory has gauge group $SU(\Nf-N)$.

As we saw earlier, the squarks of the electric theory  obtain soft masses 
through their gauge interactions with the heavy flavor $H$, ${\bar H}$.
In the magnetic theory, the higgsing is accompanied by supersymmetry breaking.
The heavy gauge multiplets corresponding to the broken generators have masses 
that are split due to the supersymmetry breaking (we give explicit expressions
for these masses in appendix~B).
The remaining squarks couple directly to these heavy gauge multiplets and 
will therefore receive soft masses starting at the one-loop order.
Another consequence of the $h$, $\hb$ and $P$ vevs is that the ``mixed'' 
mesons $V^i$ and $\Vb_i$, and the broken components of the squarks,
$q_{i}^{\N+1}$and $\qb^i_{\N+1}$ now obtain tree-level masses through the 
superpotential.
We then have additional heavy multiplets with supersymmetric mass, $\lambda v$, 
and splittings proportional to $F_h$.
The remaining squarks and scalar mesons couple to these through Yukawa 
interactions and receive soft masses, again starting at one-loop.
Thus, below the scale $v$ we have an $SU(\N)$ theory with the 
dual quarks $q_i$, $\qb^i$, and the mesons $M^i_{j}$
with  $i,j = 1\ldots\Nf$.
The scalar components of these fields acquire masses through their gauge and 
Yukawa interactions to the heavy multiplets.

To summarize, at low energies we  have constructed two theories 
related by Seiberg duality and with soft supersymmetry-breaking
terms, including scalar masses, gaugino masses 
and $A$-terms.
In the following sections, we will calculate these terms,
starting with the limit of small supersymmetry breaking.

But before doing that, a few comments regarding
the source of supersymmetry breaking are in order.
We envisage a situation in which the DSB sector has a stable minimum
at finite field vevs.
We also imagine that that the coupling  of the DSB sector to the 
$SU(N)$ and $SU(\N)$ sector does not 
dramatically alter the qualitative properties of the minimum, 
so that there is still a 
minimum with no runaway fields, 
although the actual location of 
the minimum may shift between 
the electric and magnetic theories. 
In particular the $S$ vev in the magnetic theory need not be equal to $S_0$.
Nevertheless, since
the DSB sector is only used as a source for providing
SUSY breaking,
we assume that it does not shift much in 
coupling to the two theories, (in analogy to the heat bath
in thermodynamics,) at least  for small SUSY breaking.
This leads to some requirements on the different scales in the problem.
Essentially, for given (finite) parameters of the $SU(N)$ and $SU(\N)$
theories we need to choose the scales associated with the DSB
sector to be large compared to the relevant scales of the
$SU(N)$ and $SU(\N)$ sectors.
At the same time, such a choice ensures that the 
$SU(N)$  ($SU(\N)$) theory is perturbative at the scale
$S_0$ ($v$), so that our analysis is reliable.
For example, if both $SU(N)$ and $SU(\N)$ are asymptotically free,
then for any given $\Lambda$ and $\mu$, the typical scale of the DSB sector
should be high enough, with $\Lambda, \mu < S_0 \sim S$,
so that $\bar{\Lambda} < v \sim \sqrt{\mu\, S}$.
Though it should be possible to construct  a DSB sector
(albeit not necessarily an aesthetically pleasing one)
that satisfies these requirements,
this would not significantly contribute to our investigation
and so we do not do so here.
Most of our analysis is then 
qualitatively 
equivalent to an analysis
that treats $S$ as a spurion.
{}From this point of view, one superpotential coupling,
the supersymmetric mass for $H$, $\bar{H}$, is promoted to 
a spurion field, and the resulting 
soft supersymmetry breaking induced in each theory by
the gauge and/or Yukawa interactions is then calculated.


\section{Duality with small SUSY breaking }
\label{smallbreaking}

We first consider the case of small SUSY breaking.
In this limit one may reasonably hope
that  the duality and exact results obtained by 
Seiberg~\cite{Seiberg,seibergexact}
still hold approximately, with possible corrections
higher order in SUSY 
breaking~\cite{Peskin,Hsu,Hsu2}. 

In the electric theory,
the extra flavor $H$, $\bar{H}$ get  SUSY-preserving and
SUSY-breaking masses from the $A$- and $F$-type vevs of $S$,
$S_0$ and $F_0$, respectively. 
Throughout this section, we take $\Nf$ to be smaller than, 
or close to, $3N/2$, so that the magnetic theory
is weakly coupled in the IR.
The electric theory is then asymptotically free,
and we need  $S_0 > \Lambda$
in order to be able to perform perturbative calculations.

The electric theory is in the usual ``gauge mediation''
 scenario~\cite{dnns}.
At the scale $S_0$, gauginos get masses at one loop and squarks
get masses at the two-loop order. The results are well known. At leading
order in $F_0$, they are~\cite{dnns}
\beqa
\label{electricgaugino}
\tilde{M}_g(S_0)&=& \frac{\alphae(S_0)}{4\pi} \frac{F_0}{S_0}\ , \\
\label{electricsquark}
\tilde{m}_Q^2(S_0)=\tilde{m}_{\bar{Q}}^2(S_0) &=&
\frac{N^2-1}{N} \frac{\alphae^2(S_0)}{(4\pi)^2} \left(
\frac{F_0}{S_0} \right)^2 \ .
\eeqa
The squarks obtain positive masses squared. The small SUSY
breaking case we consider here corresponds to
$\frac{F_0}{S_0^2} \ll 1$ and $\frac{\alphae}{4\pi}
\frac{F_0}{S_0} \ll \Lambda$. Because the gauge
coupling $\alpha$ is always accompanied by the loop factor
$(4\pi)^{-1}$, we will absorb $(4\pi)^{-1}$ into $\alpha$
and redefine 
\beq
\label{redefine}
\alpha \equiv \frac{g^2}{16\pi^2}
\eeq 
for convenience.
The expressions with the original definition can be easily
recovered by replacing $\alpha$ by ${\alpha}/(4\pi)$.

We now turn to the magnetic theory. We know that the duality
holds when a SUSY-preserving mass $S_0$ ($F_S=0$) is added
to $H$ and $\bar{H}$: the electric theory flows to a theory
with the same gauge group and one less flavor, and the magnetic
theory gets higgsed to a theory with
one less color and one less flavor. The resulting low energy
theories are still dual to each other. When a small SUSY
breaking $F_S$ is introduced, we expect that the vacuum
still lies close to the supersymmetric case, 
($h\bar{h} \approx -\mu S_0/\lambda$,) 
so that they connect smoothly when
$F_S \to 0$. The gauge symmetry is higgsed from $SU(\N+1)$
to $SU(\N)$ by $\langle h \rangle,\, \langle \bar{h} \rangle$.
However, as discussed in the previous section, the 
masses of the heavy gauge
supermultiplets will be split 
by nonzero $F_h$ and
$F_{\bar{h}}$ due to SUSY breaking. 
Therefore, we have gauge
messengers in the magnetic theory. In addition, some matter fields
($V^i$, $q_i^{\N+1}$) also receive masses from  
the
$h$ and $\bar{h}$ vevs,
and they give extra contributions to the soft masses of the light
fields through Yukawa couplings.

We can only calculate the soft masses of the magnetic theory if
this theory is perturbative at and below the scale 
$\langle h \rangle\equiv v$.
We will therefore assume that the Yukawa couplings are not large.
In addition, if the magnetic theory is asymptotically free, we take 
$v > \bar{\Lambda}$.
This can be acheived by choosing the typical scale of the DSB sector
to be large enough.
Note that in this case, since both the electric theory and 
the magnetic theory are asymptotically free, for any given 
$\Lambda$, $\bar{\Lambda}$, 
we can choose the scale of the DSB sector to be 
sufficiently large, so that the conditions
$S_0 > \Lambda$, $v > \bar{\Lambda}$ 
hold and both theories may be analyzed
perturbatively.
However, when the electric theory is asymptotically free
and the magnetic theory is infrared free, we need 
$S_0 > \Lambda$, $v < \bar{\Lambda}$.
These conditions can only be satisfied simultaneously
if    
$\Lambda < \bar{\Lambda}$. 
We will therefore restrict our analysis to duals with scale 
$\bar{\Lambda} > \Lambda$.    
Recall that the electric theory, with scale $\Lambda$,
has a family of duals with arbitrary $\mu$ (and therefore
arbitrary $\bar\Lambda$), that are identical 
at zero energy, but may differ at finite energies. 
We can only analyze a subset of these duals, satisfying
$v < \bar\Lambda$. 
However, our qualitative results are the same
for the entire subset. 
Therefore, they may be true in general.

In the magnetic theory, the light scalars can receive one-loop
contributions to their soft masses. 
However, these  start at 
${\cal O}(F_h^4)$ \cite{DNS,GR} whereas the two-loop contributions
start at
${\cal O}(F_h^2)$. 
Therefore, for small SUSY breaking, the two-loop
contributions dominate.
To calculate them,
we can apply the method of Giudice and Rattazzi \cite{GR}.
In this method, we first derive the wavefunction renormalization
of the light field as a function of the heavy fields' threshold,
then replace the threshold by $\sqrt{X X^{\dagger}}$, where $X$
is the field which provides the masses and SUSY breaking for
the heavy fields.
The soft mass of any light field can then be
obtained from the $\theta^2 \bar{\theta}^2$ component of the 
logarithm of the wavefunction to leading order in $F_X/X$,
without any complicated diagram calculation. In the magnetic
theory, heavy fields obtain their masses from $h$ and $\bar{h}$,
so $X=h, \, \bar{h}$ in this case\footnote{
The $(\delta h^{\N+1} - \delta \hb_{\N+1})$ field also mixes with
$\delta P$ and both receive SUSY breaking effects from $F_P$.
However, the corrections due to $F_P$ should be higher order in 
$F_h$, ($F_P \sim F_h^2/v^2$,) and are not enhanced by $N_f$ or
$\N$.
These fields do not enter at one loop either.
}.

The one-loop renormalization group equations (RGE's) of the dual
quark and meson wavefunctions, and gauge and Yukawa couplings are
\beqa
\frac{d}{dt} \ln Z_q &=& 4 c_q \am -2 d_q \ay \ ,\\
\frac{d}{dt} \ln Z_M &=& -2 d_M \ay \ ,\\
\frac{d}{dt} \am &=& 2b \, \am^2\ ,\\
\frac{d}{dt} \ay &=& \ay (-4C \am +2D \ay)\ ,
\eeqa
where $\am=g_m^2/(16\pi^2)$, $\ay= \lambda^2/(16\pi^2)$, and
$c_q= \frac{(\N+1)^2-1}{2(\N+1)} \, \left( \frac{\N^2-1}{2\N}
\right)$, $d_q= N_f +1 \, (N_f)$, $d_M=\N+1 \, (\N)$,
$b=N_f-3\N-2 \, (N_f-3\N)$, $C=2c_q$, $D=2d_q+d_M$ above
(below) the scale $v$. The method of Giudice and Rattazzi can also give
the soft masses of the light fields at low energies directly, including
renormalization running effects. We, however, will calculate
the masses at the scale $v$ and separate the discussion of
the running effects for clarity. The soft masses of the dual gauginos,
dual squarks, mesons, and the trilinear scalar coupling $A$-term 
generated at the scale $v$ are
\beqa
\label{gauginomass}
\tilde{M}_{\tilde{g}} (v) &=& -2\am \left(\frac{F_h}{v} \right),\\
\label{squarkmass}
\tilde{m}_q^2 (v)=\tilde{m}_{\bar{q}}^2(v) &=& \Biggl\{ -\am^2
 \left[ 2\,\frac{\N^2-1}{\N} + (N_f-3\N-2)\,\frac{\N^2+\N+1}{\N(\N+1)}
 \right]   \nonumber \\
& & +\am \ay \left[ -2\,\frac{(\N+1)^2-1}{\N+1} +2N_f \,
\frac{\N^2+\N+1}{\N(\N+1)} \right] \nonumber \\
& &  -\ay^2 \left[ N_f-\N-3 \right] {\Biggr\}} 
\left(\frac{F_h}{v}\right)^2 \ , \\
\label{mesonmass}
\tilde{m}_M^2 (v) &=& \Biggl\{ -\am \ay \left[ 2\,\frac{\N-1}{\N+1}
 \right]  \nonumber \\
& &  +\ay^2 \left[ 2N_f-2\N +3 \right] {\Biggr\}}
\left(\frac{F_h}{v}\right)^2 \ , \\
\label{Aterm}
A(v) &=& \Biggl\{ 2\,\frac{\N^2+\N+1}{\N(\N+1)}\, \am -3\ay \Biggr\}
\left(\frac{F_h}{v} \right)\ .
\eeqa
If we assume that  the DSB sector is not drastically altered by the 
coupling to the electric and magnetic theories,
as discussed at the end of section~\ref{framework},
we find, 
minimizing the potential of the magnetic theory, 
that for small SUSY
breaking and $S_0 \gg \mu$ (as required for perturbative calculations),
we have 
$\frac{F_h}{v} \sim \frac{1}{2} \frac{F_0}{S_0}$.
Although we will only concentrate  on the sign of the scalar
masses, this shows that the soft breaking masses generated in the
magnetic theory are about the same order as in the electric theory.

To study the low energy
theory, we have to evolve the soft breaking terms down to 
low scales. We can
do so if the couplings remain perturbative at low energies. 
Therefore, we will consider the following two cases 
in the duality regime:
the magnetically free (MF)
case ($N_f > 3\N$), and the case with a Banks-Zaks fixed point (BZ)
\cite{BZ} for the magnetic theory ($N_f/\N=3-\epsilon$ with large
$N_f$, $\N$ and $\epsilon \ll 1$). 
We can also analyze the magnetic theory for $\Nf =N +1$,
for which the electric theory confines and the magnetic theory
is completely higgsed.
In all three cases the electric theory is strongly coupled at low energies, 
so they are of special interest.

The results
in Eqs.~(\ref{squarkmass}), (\ref{mesonmass}) still depend
on the unknown couplings in the magnetic theory, especially
the relative sizes of the Yukawa coupling and the gauge coupling. 
If we work in the large $N_f$, $\N$ (and also $N_f-\N$) limit,
the results simplify to
\beqa
\label{squarkLN}
\tilde{m}_q^2 (v)=\tilde{m}_{\bar{q}}^2(v) &\approx &
- (N_f-\N) (\am-\ay)^2 \left(\frac{F_h}{v}\right)^2, \\
\label{mesonLN}
\tilde{m}_M^2 (v) &\approx & \left\{ 2(N_f-\N) \ay^2 
 -2\am \ay \right\} \left(\frac{F_h}{v}\right)^2 \ .
\eeqa
We can see that at the scale $v$ 
the dual squarks generally get negative masses squared,
and the mesons get positive masses squared (if $\ay/\am$
is not too small). This is already an interesting result. 

In the following subsections
we consider the RG running effects 
for each case separately.

\subsection{Magnetic free case} 
In the infrared theory, the couplings get weaker in running down 
toward low energies, so it is sufficient to use the one-loop
RGE's. The one-loop RGE's of the relevant quantities below $v$ are
\beqa
\label{RGalpha}
\frac{d}{dt} \am &=& (2N_f-6\N) \am^2 \ , \\
\label{RGyukawa}
\frac{d}{dt} \ay &=& \ay \left[ -4\, \frac{\N^2-1}{\N}\am 
 +(4N_f +2\N)\ay \right] \ ,\\
\label{RGMg}
\frac{d}{dt} \Mg &=& (2N_f-6\N) \am \Mg, \\
\label{RGA}
\frac{d}{dt} A &=& 4\, \frac{\N^2-1}{\N}\am \Mg +
 (4N_f +2\N)\ay A \ , \\
\label{RGmsq}
\frac{d}{dt} \msq = \frac{d}{dt} \msqb &=& -4\, \frac{\N^2-1}{\N}\am 
 \Mg^2 + 2N_f \ay (\msq +\msqb +\mm +A^2), \\
\label{RGmm}
\frac{d}{dt} \mm &=& 2\N \ay (\msq +\msqb +\mm +A^2) \ .
\eeqa
We should evolve these quantities down to the scale of the soft breaking
masses. For sufficiently small SUSY breaking, there will be enough
running for these quantities to reach their asymptotic behavior,
which we will discuss below.

The RGE's of the gauge coupling and the gaugino are easy to solve.
The solutions are
\beqa
\label{solalpha}
\frac{1}{\am(p)} &=& \frac{1}{\am(v)} + (2N_f-6\N)
 \ln \frac{v}{p} \ , \\
\label{solMg}
\frac{\Mg(p)}{\am(p)} &=& \frac{\Mg(v)}{\am(v)} \ .
\eeqa
Both $\am$ and $\Mg$ get smaller and evolve toward zero
at low energies. For the Yukawa coupling, combining 
Eqs.~(\ref{RGalpha}) and (\ref{RGyukawa}), we obtain
\beq
\label{RGyalpha}
\frac{d}{dt} \ln \left(\frac{\ay}{\am}\right)
= (4N_f+2\N)\ay - \left[ 4\, \frac{\N^2-1}{\N} +2N_f -6\N \right]
\am  \ .
\eeq
The ratio $\ay/\am$ reaches its fixed point when the right hand side
of (\ref{RGyalpha}) vanishes. So, at low energies,
\beq
\label{fixedy}
\frac{\ay}{\am} \to \frac{N_f \N -\N^2 -2}{(2N_f+\N)\N} \ ,
\eeq
and $\ay$ approaches zero too. Similarly, combining (\ref{RGMg})
and (\ref{RGA}), we obtain
\beq
\label{fixedA}
\frac{d}{dt} \ln \left( \frac{A}{\Mg} \right) =
(4N_f +2\N)\ay + 4\, \frac{\N^2-1}{\N}\am \frac{\Mg}{A} -
(2N_f-6\N)\am  \ .
\eeq
Substituting in the asymptotic ratio of $\ay/\am$, (\ref{fixedy}),
we find $\Mg/A \to 1$ in evolving to low energies.

For the scalar masses, it is convenient to define the following
two linear combinations: 
\beqa
\label{defX}
X &\equiv &\msq +\msqb +\mm \ ,\\
\label{defY}
Y &\equiv &\N \msq - N_f \mm  \ .
\eeqa
Then from (\ref{RGmsq}), (\ref{RGmm}) we have
\beqa
\label{RGX}
\frac{d}{dt} X &=& -8\, \frac{\N^2-1}{\N} \am \Mg^2 +
(4N_f+2\N)\ay X + (4N_f+2\N)\ay A^2  \ ,\\
\label{RGY}
\frac{d}{dt} Y &=& -4(\N^2-1)\am \Mg^2 \ .
\eeqa
Combining (\ref{RGX}) with (\ref{RGMg}), we obtain
\beq
\label{RGMgX}
\frac{d}{dt} \ln \left( \frac{\Mg^2}{X} \right) =
 (4N_f-12\N)\am +8\, \frac{\N^2-1}{\N} \am \frac{\Mg^2}{X}
-(4N_f+2\N)\ay - (4N_f+2\N)\ay \frac{A^2}{X} \ .
\eeq
Substituting in the asymptotic relations between $\ay$ and $\am$,
$A$ and $\Mg$, we find $\Mg^2/X \to 1$ in evolving to low
energies. Since $\Mg$ scales toward zero, we obtain the interesting
sum rule, 
\beq
\label{sumrule}
X=\msq+\msqb+\mm \to 0, 
\eeq
in the extreme infrared. 
It also tell us that some of the masses squared will be negative!
To get the individual masses we integrate (\ref{RGY}) and the result
is
\beq
\label{solY}
 (\N \msq(p)- N_f \mm(p))  =
 (\N \msq(v)- N_f \mm(v))  +
\frac{\N^2-1}{N_f-3\N} \Mg^2(v) \left[ 1-
\left(\frac{\am(p)}{\am(v)}\right)^2 \right] \ .
\eeq
Since $\am(p) \to 0$ as $p \to 0$, we can solve
for the asymptotic scalar masses from (\ref{sumrule}), (\ref{solY}),
\beq
\label{scalarmass}
\msq=\msqb=-\frac{\mm}{2}=\frac{1}{2N_f+\N} \left[ \left(
\N \msq(v) -N_f \mm(v) \right)+ \frac{\N^2-1}{N_f-3\N} \Mg^2(v) \right] \ .
\eeq
The masses in the infrared are determined by the particular
combination of the masses generated at the high scale. As
we have seen, the first part, $\N \msq(v) -N_f \mm(v)$, is
negative for large $\N$ and $N_f$. In fact, it is always
negative for $N_f > 3\N$. If there were not the gaugino mass
contribution, or if the gaugino mass contribution is small, as
for $N_f$ much larger than $3\N$, the dual squark masses squared
are negative and the meson masses squared are positive. The dual
squarks will get vevs to break the gauge and global symmetries.
We will discuss the resulting vacuum and  symmetry breaking pattern in the 
next section. For $N_f$ close to $3\N$, the gaugino mass contribution
is of the same order as the initial scalar masses, and it may
change the signs of the masses squared of the dual squarks and 
the mesons.

\subsection{Magnetic Banks-Zaks fixed point} 
For $\frac{3}{2}\N < N_f <3\N$,
there is a nontrivial fixed point for the gauge coupling
\cite{Seiberg}. The fixed point is at weak coupling in the
limit of large $\N$ and $N_f$, 
with
$N_f/\N=3-\epsilon$ fixed 
and
$\epsilon \ll 1$. To lowest order
in $\epsilon$ it can be obtained by examining the 2-loop
RGE of the gauge coupling,
\beq
\label{alpha2loop}
\frac{d}{dt}\am=2(N_f-3\N)\am^2 +2\left(-6\N^2+2\N N_f
+2N_f \frac{\N^2-1}{\N} \right)\am^3 - \frac{4N_f}{\N}
\am^2 \ay \ .
\eeq
Taking the large $\N,\, N_f$ limit and $N_f/\N=3-\epsilon$,
the fixed point occurs at
\beq
\label{fixedpt}
\am = \frac{\epsilon}{6\N} + {\cal O}(\epsilon^2)\ ,
\eeq
where we have assumed that $\ay /\am$ is ${\cal O}(1)$, which
is justified by checking the RGE of the Yukawa coupling and finding
$\ay/\am \to 2/7+{\cal O}(\epsilon)$.
At this fixed point, the perturbative expansion parameter,
$\N\am \approx \epsilon/6$, is much smaller than 1, so we
can still trust the 
perturbation theory. Most of the analysis
for the magnetic case goes through without modifications, except
for the gaugino mass part, which we now  discuss.

Similar to the gauge coupling, the one-loop $\beta$-function
coefficient is ${\cal O}(\epsilon)$ for the gaugino mass. We
have to include higher loop effects. The 2-loop RGE of the
gaugino mass in this limit is
\beq
\label{gaugino2loop}
\frac{d}{dt} \Mg = -2\epsilon \N \am \Mg + 24 \N \am^2 \Mg \ .
\eeq
When the gauge coupling approaches its fixed point, $\alpha_*
\approx \frac{\epsilon}{6\N}$, the $\beta$-function is positive
for the gaugino mass, so the gaugino mass decreases toward
0 in running down the energy scale. This can also be seen from
the exact relation between $\am$ and $\Mg$ obtained by
Hisano and Shifman \cite{HS},
\beq
\label{exact}
\frac{\am \Mg}{\beta(\am)} = 
\mbox{RG invariant,}
\eeq
and as $\am$ approaches the fixed point, $\beta(\am) \to 0$,
hence $\Mg \to 0$ too~\footnote{Note that this result implies 
that the gaugino mass vanishes at {\it any} IR fixed point.}. 
Then following the analysis 
of the previous subsection, 
we have  similar relations in the extreme
infrared,
\beqa
\label{sumruleBZ}
X &=& \msq+\msqb+\mm \to 0, \\
\label{solYBZ}
 (\N \msq(p)- N_f \mm(p))  &=&
 (\N \msq(v)- N_f \mm(v))  + M_R^2 \ ,
\eeqa
where
\beq
\N \msq(v)- N_f \mm(v) \approx -2\N^2 \am^2 +4\N^2 \am \ay
-14 \N^2 \ay^2
\eeq
in this limit, and
\beq
M_R^2 \equiv \int^{\ln v}_{-\infty} 4(\N^2-1)\am \Mg^2 dt
\eeq
is the gaugino mass contribution. We can calculate $M_R^2$
using (\ref{exact}) and (\ref{alpha2loop}) in the BZ limit.
Denoting $\am(v)\equiv \alpha_0$ and $\alpha(0)
\equiv \alpha_* \approx \frac{\epsilon}{6\N}$,
\beqa
\label{MR}
M_R^2 &=& 4(\N^2-1) \frac{\ao^2 \Mg^2(v)}{\beta^2(\ao)}
   \int_{\af}^{\ao} \frac{\beta(\am)}{\am} d\am \nonumber \\
&\approx &  4(\N^2-1) \frac{\ao^2 \Mg^2(v)}{\beta^2(\ao)}
   \int_{\af}^{\ao} (-2\epsilon \N \am +12 \N^2 \am^2) d\am \nonumber \\
&=& 4(\N^2-1) \frac{\ao^2 \Mg^2(v)}{\beta^2(\ao)}
\left[ (-\epsilon \N \ao^2 + 4\N^2\ao^3)-(-\epsilon \N \af^2 + 4\N^2\af^3)
\right] \nonumber \\
& \approx & (\N^2-1)\, \ao^2 \Mg^2(v) \,\frac{12\N}{\epsilon^3}\,
\frac{\left(1-3\frac{\ao^2}{\af^2}+2\frac{\ao^3}{\af^3}\right)}{\left(
\frac{\ao^2}{\af^2}-\frac{\ao^3}{\af^3} \right)^2} \nonumber \\
& \approx & (\N^2-1)\,\ao^2 \Mg^2(v) \,\frac{\N}{18\N^3 \ao^3}\,
\frac{\left(2-3\frac{\af}{\ao}+\frac{\af^3}{\ao^3}\right)}{ \left(
1-\frac{\af}{\ao} \right)^2} \ .
\eeqa
This gaugino mass contribution is generically enhanced by $\N$ compared
with the $\N \msq(v)- N_f \mm(v)$ part if $\ao <\af$ or
$\N \ao \ll 1$, and $\ay$ is not much larger than $\ao$.
Hence it could make the dual squark masses squared 
positive,
and the meson masses squared 
negative, in evolving to low energies.

\subsection{The completely higgsed magnetic theory ($N_f=N+1$)}
In this subsection we consider the case
$N_f=N+1$ in the electric theory. Without SUSY breaking,
the low energy theory is confining without chiral symmetry
breaking~\cite{seibergexact}. 
The low energy degrees of freedom are described by
the baryons $B, \, \bar{B}$ and the mesons $M$, with the
effective superpotential~\cite{seibergexact}
\beq
\label{Weff}
W_{\rm{eff}}= \frac{1}{\Lambda_L^{2N-1}} (M^i_j B_i \bar{B}^j
- \det M)\ .
\eeq
It is interesting to see what SUSY breaking 
masses 
baryons and mesons
receive when SUSY breaking masses are added for the elementary
squarks. We follow the same procedure as before. Dualizing
the electric theory with one more flavor (which receives 
SUSY-preserving and SUSY-breaking masses from $S$), the magnetic
theory has an $SU(2)$ gauge symmetry, which is then broken
completely by the $h, \, \bar{h}$ vevs. The low energy dual
quarks 
correspond to the baryons of the electric theory~\cite{Seiberg}.
The $\det M$ term in the superpotential is generated by
instantons~\cite{Seiberg}.
We again use the method of Giudice and Rattazzi to calculate 
the masses in the magnetic theory, which should be sufficient
for $N_f>4$, 
for which 
the $\det M$ term is nonrenormalizable and
its effect is probably small. The masses of the dual squarks
(baryons) and the mesons at the scale $v$ are
\beqa
\label{bmass}
\msq(v)= \msqb(v) &=& \left(-\frac{3N_f-15}{2} \am^2 +(3N_f-3)
 \am \ay - (N_f-4) \ay^2 \right)\,
\left( {F_h\over v} \right)^2
\ , \\
\label{mmass}
\mm(v) &=& (2N_f+1) \ay^2 \, \left( {F_h\over v} \right)^2
\ .
\eeqa
After running to low energies, we have in the extreme infrared,
\beqa
\label{sumrule1}
\msq +\msqb +\mm &=& 0 \ ,\\
\label{diff1}
\left. \msq(p)- N_f \mm(p) \right|_{p \to 0}
&=& \msq(v)- N_f \mm(v).
\eeqa
Substituting in (\ref{bmass}) and (\ref{mmass}), we obtain
\beq
\label{confinement}
\msq = \msqb = -\frac{\mm}{2}= \frac{1}{2N_f+1} \left(
- \frac{3N_f-15}{2} \am^2 +(3N_f-3)
 \am \ay - (2N_f^2+2N_f-4) \ay^2 \right).
\eeq
For small $N_f$, the result depends on the relative strength of
$\am$ and $\ay$ which we do not know. However, for large
enough $N_f$, the dual squark (baryon) masses squared are negative, 
and  baryon number is  broken. 
This may have interesting
implications 
in compositeness models based  on 
$SU(N)$ with $\Nf = N + 1$  or analogous theories.
For $N=2, \, N_f=3$, we do not know how to calculate
the masses because the instanton generated superpotential
becomes a renormalizable Yukawa interaction. However, there
is no distinction between the baryons and the mesons in this case. 
Therefore,
if they still satisfy the sum rule (\ref{sumrule1}), their masses will
vanish to leading order\footnote{Similar spectra of soft breaking 
masses in the dual and confining theories are also obtained by a
different method \cite{nima}.}.



\section{Vacua and symmetry breaking patterns}
\label{vacuum}

As we saw in the previous section, 
some of the fields in the magnetic theory 
acquire negative masses squared. 
As a result, some of these fields will develop
vevs and break the gauge and/or global symmetries. 
In this section, we
discuss the vacua and the symmetry breaking patterns 
for the two possibilities we have encountered:
$\msq < 0$, $\mm > 0$ and $\msq > 0$, $\mm < 0$.

We start with  negative  dual-squark masses squared 
and positive meson masses squared,  
$\msq < 0$, $\mm > 0$.
For small SUSY breaking,
this happens in our examples of 
the magnetic free theories (with $N_f \gg 3\N$,
{\it i.e.,} $N+1 < N_f \ll \frac{3}{2} N$ in the electric theory),
and in the duals of the 
$\Nf = N + 1$ theories
with large enough $N_f$.

After integrating out the heavy fields,
the potential in the low energy theory is
\beqa
\label{lowpotential}
V \ &=&\ 
\lambda^2\,\left( 
{\left\vert \, M^i_{j} \, q_i \right\vert}^2 \ + \ 
{\left\vert \, M^i_{j}\,  \qb^j \right\vert}^2 \ + \ 
{\left\vert\, q_i\cdot \qb^j \,\right\vert}^2 \right) 
  \ + \
 \msq\, (\vert q_i\vert^2 \ +\  \vert \qb^j\vert^2 ) 
\\~\nonumber
\ &+& \
\mm\, \vert M^i_j\vert^2 
\ +\  {g_m^2\over 2}\, D^2  \ +
\mbox{$A$-terms} \ .
\eeqa
For $\msq < 0$, the potential is unbounded from below
along ``runaway"  directions which correspond to the 
baryon directions\footnote{In Ref.~\cite{DHoker}, 
where the full theory is described by a potential 
similar to~(\ref{lowpotential}),
the case that
all dual squarks have negative masses squared is not considered
because of the runaway behavior.
In our case however, physics at the scale $v$ stabilizes 
the runaway.}. 
Up to symmetry rotations and exchange of
$q$ and $\bar{q}$, the runaway directions take the following form,
\beq
\label{sqvev}
q = \left( \begin{array}{cc}
             \begin{array}{cc}
               \begin{array}{cc} u & {}\\ {} & u
               \end{array} & \mbox{\LARGE 0} \\
               \mbox{\LARGE 0} &
               \begin{array}{cc} \ddots & {}\\ {} & u
               \end{array}
             \end{array}
             & \mbox{\Huge 0}
           \end{array} \right), \hspace{0.3in}
\bar{q}=0\ .
\eeq
The magnetic gauge group is completely higgsed, and the global symmetry
is broken to $SU(\N)_L \times SU(N_f-\N)_L \times SU(N_f)_R \times
U(1)'$, (or $SU(N_f-1)_L \times SU(N_f)_R \times U(1)'$ for the 
case $N_f=N+1$ 
with large enough $N_f$,) where $U(1)'$ is a linear combination of
 $U(1)_B$ and a
$U(1)$ subgroup of $SU(N_f)_L$.

In the full magnetic theory, the
runaway direction~(\ref{sqvev})
is stabilized due to the presence of the
heavy fields.  
Recall that the fields $h$ and $\hb$ had vevs of the form
$\langle h^{\N+1} \rangle
= - \langle \bar{h}_{\N+1} \rangle =v$.
Hence in the full theory, the direction~(\ref{sqvev}) 
is not  D-flat.
Instead, the potential has a minimum with the  $h$ and $\bar{h}$ 
vevs slightly shifted from their original values. 
Then of course 
our calculation of 
the light field masses may no longer be valid, and 
these 
masses will depend on the shifts. 
We will  ignore this dependence in the calculation,
assuming the $h$, ${\bar h}$ shifts are small.
This will prove to be a self-consistent assumption.

Since some of the heavy fields are now relevant,
we need to re-examine the full potential~(\ref{potential}).  
Extremizing the potential with respect
to $P$ we obtain (Eq.~(\ref{peq}))
\beq
\label{peq1}
P = \frac{\mu F_{SB}}{\lambda^2 (|h|^2 +|\bar{h}|^2) +\mu^2}\ ,
\eeq
where $F_{SB} \equiv - \frac{\partial W_{SB}}{\partial S}$.
Letting $\langle h^{\N+1} \rangle =v+\delta \ , \,
\langle \bar{h}_{\N+1} \rangle =-v+\bar{\delta}$ and substituting
(\ref{peq1}) into the potential, we have
\beq
\label{Vtree}
V_{\rm tree}= \lambda^2 \left[ -v^2 -v(\delta -\bar{\delta})
+\delta \bar{\delta} +\mu S \right]^2 -
\frac{\mu^2 F_{SB}^2}{\lambda^2 \left[ 2v^2 +2v(\delta -\bar{\delta})
+\delta^2 +\bar{\delta}^2 \right]+\mu^2}\ .
\eeq
In addition, there are contributions to the potential coming from
D-terms and from soft SUSY breaking squark masses,
\beqa
\label{VD}
V_D &=& \frac{1}{2} g_m^2 \left| \frac{1}{\sqrt{2\N (\N+1)}}
\left[ \N (v+\delta)^2 -\N u^2 - \N (-v+\bar{\delta})^2 \right]
\right|^2 \nonumber \\
&=& \frac{\N}{4(\N+1)} g_m^2 \left[ 2v(\delta+\bar{\delta})
 +(\delta^2 - \bar{\delta}^2) -u^2 \right]^2 \ , \\
\label{Vsoft}
V_{\rm soft} &=& -\N m_1^2 u^2\ ,
\eeqa
where $m_1^2 \equiv -\msq >0$, and we will also ignore its dependence
on the scale $u$ to a first approximation. 
Minimizing the potential~(\ref{Vtree}) we
find $\bar{\delta} \approx \delta$ to leading order.
Expanding (\ref{Vtree}) to the lowest order in $\delta$
and simplifying it 
using
Eq.~(\ref{min}) and $\bar{\delta}=\delta$,
we find
\beq
\label{Vtree1}
V_{\rm tree} \approx V_{\rm tree}(\delta=0) + \frac{4\lambda^2
 \mu^2 F_{SB}^2}{(2\lambda^2 v^2 +\mu^2)^2}\, \delta^2\ .
\eeq
Combining this with $V_D$, $V_{\rm soft}$, and using the relation
(\ref{fvevs}), $\frac{F_h}{v}=\frac{\lambda \mu F_{SB}}{2\lambda^2 v^2
+\mu^2}$, the total potential is given by
\beq
\label{Vtotal}
V= V(u=0, \delta=0) + \frac{\N}{4(\N+1)} g_m^2 (4v\delta-u^2)^2
 - \N m_1^2 u^2 +4 \left( \frac{F_h}{v} \right)^2 \delta^2\ .
\eeq
Now we can minimize the potential with respect to $u$ and $\delta$.
We find that the minimum occurs at 
\beq
\label{delta}
\delta = \frac{\N m_1^2}{2 \left(\frac{F_h}{v} \right)^2} \, v\ ,
\eeq
and
\beq
\label{u}
u \approx 2 \protect\sqrt{v\delta}\ .
\eeq
Because $m_1^2 \sim \N \alpha_m^2 (\frac{F_h}{v})^2$, $\delta /v$
is suppressed by the loop factor $(\N \alpha_m)^2$. This justifies
our assumption that $\delta$ is small compared with $v$.

Below the symmetry breaking scale $u$, the remaining light fields
are $q^a_{i>\N}$, the phase of $u$, which correspond to the
Goldstone fields of the broken global symmetry, and $M_j^{i>\N}$.
If duality is still good 
for small SUSY breaking, 
the electric theory may have the same symmetry breaking pattern.
The symmetry breaking scale could be stabilized by strong dynamics
or by a nonminimal K\"{a}hler potential, and the most natural scale
will be the strong coupling scale $\Lambda$ of the electric 
theory\footnote{Note that in the magnetic theory,
$\delta$ and $u$ are independent of the SUSY breaking 
to leading order, so the symmetry
breaking scale does not get smaller as SUSY breaking decreases.
If this is also true in the electric theory, the only natural
scale for symmetry breaking is $\Lambda$.}.

Finally, we note that there is no minimum
with the squarks and anti-squarks developing equal vevs
with the symmetry broken to 
$SU(\N)_D\times SU(\Nf-\N)_L\times SU(\Nf-\N)_R \times
U(1)$. 
Such directions are D-flat even in the full theory,
and may be studied using the potential~(\ref{lowpotential}).
However, the only stationary points along these directions are
saddle points, and the theory slides towards the ``runaway" baryonic
directions discussed above.
Interestingly, the result that only $SU(N_f)_L$ or $SU(N_f)_R$
is broken has some similarity with the result obtained for
$SU(2),\, N_f=2$ with small soft breaking in Ref.~\cite{A-G}.

We next turn to the case with positive dual squark masses squared
and negative meson masses squared,
$\msq > 0$, $\mm < 0$. 
This can happen 
when
the gaugino mass contribution changes the signs of the dual squark and 
the meson masses squared through RG evolution,
as 
could happen for the 
magnetic Banks-Zaks fixed point discussed in the previous 
section\footnote{
Whether the gaugino mass contribution alters the signs of the scalar
masses squared also depends on the relative sizes of the gaugino mass
and the scalar masses, in addition to $\N_f$ and $\N$. In the examples 
we studied, the relation between the gaugino and the scalar masses
is fixed because a 
specific messenger sector was
chosen. 
Different ratios of the gaugino and the scalar masses may be
obtained from different messenger sectors. For example, increasing
the number of messenger flavors will increase the ratio of gaugino
to scalar masses in both the electric and magnetic theories.}.
It is also
the case in the dual theory of an electric theory with negative squark
masses squared, as described in Appendix~\ref{appenA}.

In the low energy theory, there are also runaway directions which 
correspond to the meson directions. 
Classically, $M$ can run away along the directions
\beq
\label{classicalrunaway}
M = \left(    \begin{array}{cc}
               \begin{array}{cc} r_1 & {}\\ {} & r_2
               \end{array} & \mbox{\LARGE 0} \\
               \mbox{\LARGE 0} &
               \begin{array}{cc} \ddots & {}\\ {} & r_{N_f}
               \end{array}
             \end{array}
              \right) \ .
\eeq
However, in the supersymmetric case, when
${\rm rank}(M) > N_f-\N$, a nonperturbative superpotential is
 generated~\cite{Seiberg}. 
In the presence of small SUSY breaking,
this superpotential  still describes the nonperturbative effects
 at  lowest order~\cite{Hsu}.
For example, if all $r_i$ are equal
($=r$), the nonperturbative superpotential will be 
$\sim r^{\frac{N_f}{\N}}$, resulting in a potential 
$\sim r^{2(\frac{N_f}{\N}-1)}$. For $N_f> 2\N$, (true in 
the cases we considered,) the potential has a
minimum along the direction of $r$ at some nonzero $r$.
This is only a saddle point, however, 
which
is unstable
in the directions where some $r_i$ increase and  others decrease.
In fact, for ${\rm rank}(M) > N_f-\N$, the directions in which
larger $r_i$ increase and smaller $r_i$ decrease are always
unstable when the nonperturbative superpotential is included,
so some smaller $r_i$ will slide toward zero.

The vacuum will
then
 run away along the directions
\beq
\label{mvev}
M = \left( \begin{array}{cc}
             \begin{array}{cc}
               \begin{array}{cc} r_1 & {}\\ {} & r_2
               \end{array} & \mbox{\LARGE 0} \\
               \mbox{\LARGE 0} &
               \begin{array}{cc} \ddots & {}\\ {} & r_{N_f-\N}
               \end{array}
             \end{array}
             & \mbox{\Huge 0} \\
             \mbox{\Huge 0} & \mbox{\Huge 0}
           \end{array} \right) \ .
\eeq
There are $\N$ flavors of dual quarks left after integrating out
the heavy flavors which get masses from $r_i$. Then, the light
dual quarks will confine into ``dual mesons'' $N^l_k, \,
k,\, l= N_f-\N+1, \ldots, N_f$, and 
baryons $b',\, \bar{b}'$, with a quantum modified constraint
\beq
\label{constraint}
\det N - \bar{b}' b' = \bar{\Lambda}_L^{2\N} \ .
\eeq
The Yukawa interaction $\lambda M^k_l q_k \qb^l, \,\,
k,\, l= N_f-\N+1, \ldots, N_f$, turns into a mass term,
\beq
\label{yukawamass}
\lambda M^k_l N^l_k \ , \hspace{0.3in}  k,\, l= N_f-\N+1, \ldots, N_f \ ,
\eeq
which forces $M^k_l, \, N^l_k, \,\, k,\, l= N_f-\N+1, \ldots, N_f$,
to be zero. The baryons $b',\, \bar{b}'$ will get vevs from
Eq.~(\ref{constraint}) and break  
$U(1)_B$\footnote{As long as the
baryons are fixed by the constraint, $\bar{b}' b' =-\bar{\Lambda}_L^{2\N}$,
they do not represent new independent fields along the direction
of varying $\bar{\Lambda}_L$, or equivalently, $r_i$. Therefore,
at least to  lowest order,
there should not be  extra independent soft breaking masses
and potential along that direction for the baryons.}.
The global symmetry
is broken, if all $r_i$ are equal, to $SU(N_f-\N)_V \times
SU(\N)_L \times SU(\N)_R \times U(1)'$, where $U(1)'$ is a linear
combination of $U(1)_B$ and $U(1)$ subgroups in $SU(N_f)_L$
and $SU(N_f)_R$. Interestingly, the symmetry breaking patterns
of the runaway vacua exactly correspond to what we expect in an
electric theory with negative squark masses squared 
(see Appendix~\ref{appenA}). 
This gives us some confidence in our analysis.

For the theories in which the meson masses squared are positive
at high scales, but only turn negative at low energies through 
the RG evolution, {\it e.g.}, the magnetic BZ theory discussed
in section~\ref{smallbreaking}, the mesons can not run away in the
full theory because their masses squared are positive at high scales.
Their vevs $r_i$ will be stabilized at the scale where their
masses squared turn negative after including the one-loop
effective potential~\cite{CW}. Again, we should not
trust the symmetry breaking scale since it is large. The natural
scale for symmetry breaking in the electric theory is the strong 
scale $\Lambda$.


\section{The large SUSY breaking limit}  %
\label{largebreaking}

So far we have studied the vacuum of the magnetic
theory when this theory is weakly coupled in the IR,
and in the presence of very small supersymmetry breaking,
such that the soft masses in the electric theory are small
compared with its strong coupling scale.
While we could analyze the behavior of the magnetic
theory at low energies, we had no direct information about the low
energy behavior of the electric theory, so we
could not  compare the two.

We now turn to the other limiting case,
that of large supersymmetry breaking.
Here, the soft masses generated in each theory are large
compared with the scale of the theory (assuming it is asymptotically free).
For finite $\mu$, this can be ensured by taking the typical scale of
 the DSB sector to be much larger than $\Lambda$, $\mu$.
 This limit is interesting for two reasons.
First, and most obvious, the electric theory at low energies
approaches non-supersymmetric QCD.
Second, as we will see, we will be able to confront our findings
for the chiral symmetry breaking in the magnetic theory
with some known results for the electric theory.
In many cases, it will be possible to rule out the proposed
duality based on this comparison.

The soft masses in the electric theory are given 
in Eqs.~(\ref{electricgaugino}), (\ref{electricsquark}).
When the electric theory is asymptotically free,
we take $\Lambda \ll \alpha_e {F_0\over S_0}$. 
The squarks and gluinos are heavy compared with the scale of 
$SU(N)$, and the low energy theory is QCD, with $N$
colors and $N_f$ fermion flavors.
For sufficiently small $N_f$, 
the chiral symmetry of the theory is expected to be broken to 
$SU(\Nf)\times U(1)_B$. 
This is experimentally known for ``real" QCD, and was shown to hold
for large $N$ with $\Nf$ fixed~\footnote{More recent work claims this 
is the case for $\Nf < 4N$~\cite{Terning}.}.
For larger values  of $\Nf$, with fixed $N$, the behavior of the theory 
at low energies is not known.
In particular, 
it is possible that  some region of $\Nf$, $N$ exists,
where the theory flows to an infrared fixed point with unbroken
 chiral symmetry, 
as found by Seiberg for the supersymmetric case~\cite{Seiberg}.
We do know however, that the  vector symmetry of the theory,
$SU(\Nf)_V\times U(1)_B$ remains unbroken~\cite{VW}.
As $\Nf$ is increased beyond 11$N$/2, the theory becomes
IR free, and the full global symmetry group remains 
unbroken at low energies.

The soft masses in the magnetic theory, 
to leading order in $F_h$, are given by the two-loop 
contributions~(\ref{squarkmass}), (\ref{mesonmass}).
They are of the order $\sqrt{\N} \alpha\, f\, v$
with $f\equiv F_h/v^2$, and $\alpha$ corresponds to either the gauge
or the Yukawa coupling constant.
When this theory is asymptotically free, we
take $\bar{\Lambda} \ll \sqrt{\N} \alpha\, f\, v$.
This can be achieved by taking $\mu^\prime$ to be sufficiently small.

As discussed in section~\ref{smallbreaking}, these masses also receive one-loop
contributions which are of the order
${1\over 4\pi} f^2\, v$, to leading order in $f$.
(For some details of the calculation and the explicit one-loop expressions,
see  appendix~\ref{oneloop}).
The two-loop contributions dominate for $f \ll \sqrt{\N} \alpha$,
and the one-loop contributions are more important for 
$f > \sqrt{\N} \alpha$. For the large SUSY breaking mass limit,
there are two possibilities: We can either increase both $F_h$
and $v$ while keeping $f$ small so that two-loop contributions still
dominate, or increase $F_h$ only, so that at some point the one-loop
contributions become more important. We will consider both
possibilities in the following.

We first consider the region where the two-loop contributions
dominate.
As before, we are mainly interested in the sign of the masses.
Unlike in the case of small supersymmetry breaking, 
which we considered before, the scale of the soft masses 
is not very different from $v$ and we neglect running effects.
The squark and scalar meson
masses~(\ref{squarkmass}) and~(\ref{mesonmass})
are  functions of $\Nf$, $\N  = \Nf-N$,
$\am$ and $\alpha_\lambda$. 
In the limit $\Nf, N \gg 1$ with either 
 $\Nf - N $ fixed or $\Nf/ N$ fixed we find generally
$\msq  < 0$ and $\mm > 0$ ($\mm$ may change sign if 
$\am/\alpha_\lambda > N$).
As discussed in section~\ref{vacuum}, either the scalar quarks or the
scalar antiquarks then develop vevs along a baryonic direction,
and the gauge symmetry is completely higgsed.
The theory then has a stable minimum with the chiral
symmetry broken to $SU(\Nf)\times SU(\Nf-\N)\times SU(\N) \times U(1)'$.
However, the electric theory is $SU(N)$ with $\Nf$ flavors of
fermion fields, where we do not expect vector-like symmetries to be 
broken~\cite{VW}.
Thus, the magnetic theory we consider does not seem to give
a valid dual description of the electric theory in the infrared
in this case.

We now go on to discuss the one-loop contributions
to the soft masses, which dominate when $f$
is large compared to the gauge and Yukawa couplings.
The explicit one-loop expressions are given in appendix~\ref{oneloop}.
In this case,
the scalar meson mass squared~(\ref{mesononeloop})
is always negative.
The sign of 
the squark mass squared, $\tilde{M}^2_q$ 
 (Eqs.~(\ref{oneloopsq})-(\ref{mprime})),
depends on $\N$, $f/\lambda$, and $g_m/\lambda$.
We will therefore only give a qualitative description of the behavior
of this sign, keeping the Yukawa coupling $\lambda$ fixed.
Roughly speaking, the Yukawa contribution is positive and the gauge 
contribution is negative, so that 
$\tilde{M}^2_q$ is large and positive for small $g_m$ and decreases as
$g_m$ increases.
For large values of $\N$, say $\N > 50$, 
$\tilde{M}^2_q$ is positive for most relevant values of $g_m$ and $f$
(We have varied $g_m/\lambda$ between 0.05 and 200, and $f/\lambda$ between
0.1 and 0.95.).
For lower values of $\N$, 
$\tilde{M}^2_q$ becomes negative below a certain value of $f$.
For example, for $\N =10$, we get a negative mass squared
for $g_m/\lambda > 0.05$ and $f/\lambda < 0.6$. 
For $\N=3$, $\tilde{M}^2_q < 0$ in the entire region
$0.05 <g_m/\lambda < 200$, $ 0.1 < f/\lambda < 0.95$.

To summarize, at one-loop, the scalar meson masses
squared are always negative, and the sign of the squark mass 
squared can be either positive or negative.
For large $\N$, it is almost always positive, and for $\N=3$
it is almost always negative. 
For intermediate $\N$, it is positive for large $f$, and
becomes negative  as $f$ is decreased.

We immediately see that there is no region where the
full chiral symmetry remains unbroken.
Thus, for large $\Nf$, such that the electric theory
is IR free, the two theories are clearly different in the IR.
Furthermore, the magnetic theory can not correspond to an
IR fixed point with the full chiral symmetry unbroken.

For  small $\N$, 
the global symmetry of the electric theory
at low energies is known to be $SU(\Nf)_V\times U(1)_B$,
while in the magnetic theory, 
the squark mass squared is almost always
negative. 
If, as a result, some squarks develop vevs, 
we get  different patterns of global symmetry breaking
in the two theories.
It is still possible however that only the mesons develop
vevs, thus generating positive squark masses 
and driving the squarks to the origin.
If the meson vevs are all equal, we may
then obtain the same pattern of symmetry breaking as in the electric theory.
This is also possible when $\mm  < 0$ and $\msq > 0$,
as is the case for intermediate values of $\Nf$ 
with  $f$ sufficiently large. 
However, 
there is no tree-level term
to stabilize the potential  
along this direction.
The meson vevs may  be stabilized by a nonminimal
K\"ahler potential and/or  nonperturbative effects,
but because SUSY is badly broken, we do not know what
the nonperturbative potential is  and there is no small expansion
parameter proportional to the SUSY breaking for the K\"ahler
potential. 
Furthermore, nonperturbative effects will presumably be relevant
only for meson vevs larger than the soft masses,
which are not much smaller than the scale $v$ in the case at 
hand.
Hence, we cannot determine whether a stable minimum
exists with the symmetry broken to $SU(\Nf)_V\times U(1)_B$.
In fact, if we estimate the nonperturbative potential by
$\sim \Lambda_L^4(M)$, where $\Lambda_L(M)$ is the strong coupling
scale after integrating out the quark fields which obtain masses from
the meson vevs, then much like in the case discussed in the previous
section, the potential is lifted at large scales along the
direction $M \propto I$ for a certain range of $N_f$, but the vacuum
will slide away, with  meson fields obtaining
different vevs.
Some of the vector 
symmetries of the theory will then be broken, in contradiction to 
what we expect for the electric theory.

\section{Discussion and Conclusions }
\label{conclusions}

In this work, we studied the infrared behavior of theories
related by Seiberg duality in the presence of supersymmetry
breaking. The difficulty of not knowing what the soft SUSY
breaking terms in one theory map into in its dual is overcome by
generating the soft breaking terms in both theories from
the same SUSY breaking source, {\it i.e.}, by coupling them
to the same sector which breaks supersymmetry spontaneously.
In the duality of ${\cal N}=1$ SUSY QCD theories, giving a
mass term to a flavor of quark superfields 
corresponds to higgsing the dual gauge group to
a smaller one. 
Similarly, as we saw here
in the presence of SUSY breaking,
generating soft breaking masses in the electric theory by
heavy {\it matter} messengers
corresponds to generating
soft breaking masses in the magnetic theory by heavy {\it gauge}
messengers. 
Assuming a canonical K\"ahler potential,   
we found that the soft breaking scalar masses
squared generated in the magnetic theory are often negative,
leading to symmetry breaking in the magnetic theory.

If duality still holds approximately for small SUSY breaking
masses (much smaller than the strong coupling scale) and our
analysis is valid, it may be used for studying  strongly
coupled SUSY QCD with small SUSY breaking masses. 
In particular,
for a range of $N+1 \leq N_f < N_f^0$ in the electric theory,
where $N_f^0$ is close to but somewhat larger than
$\frac{3}{2} N$, the dual magnetic theory is weakly coupled and we 
can analyze it. 
Our results can be roughly summarized as follows:
We obtain an interesting sum rule,
$\msq+\msqb+\mm =0$ in the deep infrared, which means that either the 
masses squared of the dual squarks or the mesons are negative.
Whether $\msq < 0$ or $\mm < 0$ 
depends on the relative sizes of a certain combination of 
the scalar masses
at the high scale where they are induced,
and the RG contribution from the gaugino mass. 
These in turn depend on $N$, $\Nf$, and the ratio of
Yukawa to gauge coupling.
In the region $N_f \ll 3N/2$, the 
magnetic theory is very weakly coupled 
at low energies and hence the gaugino contribution
is small, 
and we find that  $\msq<0, \, \mm>0$ in the deep infrared.
The theory has a stable minimum
with the symmetry broken to
$SU(N_f-N)_L \times SU(N)_L \times SU(N_f)_R \times U(1)'$,
or with $L$ and $R$ exchanged.
When the gaugino mass contribution is large, which could happen
because of a larger gaugino mass or a stronger gauge coupling at
low energies in the magnetic theory, the signs of the scalar
masses squared may be altered due to the RG contribution
from the gaugino mass, so that $\msq>0, \, \mm<0$ in the
infrared. Including the nonperturbative effects in 
SQCD~\cite{Seiberg, seibergexact}, we find that the symmetry
is then broken to $SU(N)_V \times SU(N_f-N)_L \times SU(N_f-N)_R \times U(1)'$.

The relative size of the gaugino 
and squark mass is fixed in our models because we have only
considered 
a specific
messenger sector.
However,
 different ratios of 
gaugino to scalar masses can be obtained from different
messenger sectors, ({\it e.g.}, more flavors of the messenger fields).
Because the sign of the scalar mass  depends on the gaugino mass
and on the strength of the gauge coupling in the magnetic theory
at low energies, we expect the following behavior.
For fixed $N, \, N_f$, with $N_f/N$ within the range we studied, 
there will be a critical $\Mg^c$
such that for $\Mg < \Mg^c$,
the dual squark masses squared are negative, 
resulting in the symmetry breaking pattern
$SU(N_f-N)_L \times SU(N)_L \times SU(N_f)_R \times U(1)'$ 
(up to exchanging $L$ and $R$), and for $\Mg > \Mg^c$, 
the meson masses squared are negative and the symmetry
is broken to $SU(N)_V \times
SU(N_f-N)_L \times SU(N_f-N)_R \times U(1)'$.
We can also turn this statement around. For fixed $\Mg$ and
$N$, there will be a critical $N_f^c$ such that
the symmetry is broken to 
$SU(N_f-N)_L \times SU(N)_L \times SU(N_f)_R \times U(1)'$
for $N_f < N_f^c$, 
(where 
the magnetic theory is more weakly
coupled at low energies,) and 
to $SU(N)_V \times SU(N_f-N)_L \times SU(N_f-N)_R \times U(1)'$
for $N_f > N_f^c$,
(with $\Nf/N$ within the range we studied).
$N_f^c$ will depend on the relative sizes of the gaugino
and the scalar masses in such a way that it decreases as
the gaugino mass increases.

We also consider the large SUSY breaking limit. 
Below the soft SUSY breaking mass scale the squarks and gluino 
in the electric theory decouple. 
The theory becomes
ordinary nonsupersymmetric QCD
for which there are some known  results.
We can then compare them with the results we
obtain for the magnetic theory. 
In the magnetic theory we typically find that either the mesons
or the squarks or both obtain negative masses
squared, depending on the values of $\Nf$, $N$
and the gauge and Yukawa couplings.
As a result,
the magnetic theory has
no stable minimum with unbroken vector-like symmetries
within the minimal
framework we assumed.
This is in contradiction to what we expect for non-supersymmetric QCD.
The candidate duals we considered therefore do not describe
the same low-energy physics as ordinary QCD.

Throughout our analysis we have assumed a minimal 
K\"ahler potential in the dual theory.
The K\"ahler potential may contain higher-dimension
terms inversely proportional to $\Lambda$, the strong coupling scale of
the electric theory.
These terms may not be neglected, especially when SUSY breaking
is large.
Our results also depend on the Yukawa coupling of the magnetic theory,
and particularly on the assumption that it remains perturbative
at the scale where we perform calculations.
Still, the fact that these simple duals fail for large supersymmetry breaking
 is intriguing.
We do not know whether including 
a more complicated K\"ahler potential or nonperturbative
effects would modify this result,  
because there is
no systematic way of studying them. 
It is possible that the theory undergoes a phase transition
after supersymmetry is broken, and duality breaks down.

\section*{Acknowledgements}

We
would like to thank N.~Arkani-Hamed, 
N.~Seiberg, and S.~Trivedi for discussions. 
This work was
supported by the U.S. Department of Energy under the contract
DE-AC02-76CH0300.

\appendix

\section{Soft SUSY breaking masses from matter messengers}
\label{appenA}

In this appendix we consider the case where we have only
matter messengers in the magnetic theory, {\it i.e.}, $h, \, \bar{h}$
do not get vevs but receive a large mass from the vev of $P$, 
and $F_P/P^2 \ll 1$. This can be a dual description of an electric
theory which has gauge messengers, and therefore negative masses
squared for squarks~\cite{GR}. Again, we can use the method
of Ref.~\cite{GR} to calculate the 2-loop masses for the dual squarks 
and the mesons in the magnetic theory, and the results are
\beqa
\msq(P) = \msqb(P) &=& \left[ \am^2 \frac{\N'^2-1}{\N'}
 - 2\am \ay \frac{\N'^2-1}{\N'} + \ay^2 (\N'+2) \right]
 \left( \frac{F_P}{P} \right)^2 \nonumber \\
&=& \left[ \frac{\N'^2-1}{\N'} (\am -\ay)^2 + (2 +\frac{1}{\N'}) \ay^2
 \right] \left( \frac{F_P}{P} \right)^2 \ , \\
\mm(P) &=& -2\N' \ay^2 \left( \frac{F_P}{P} \right)^2 \ ,
\eeqa
where the magnetic gauge group is $SU(\N')$.
We see that the dual squark masses squared are positive and the meson
masses squared are negative.  The RG contribution from the
gaugino mass will not alter these signs in evolving to low energies.

As we saw in section~\ref{framework}, the magnetic theory we discussed
throughout the paper (corresponding to an electric theory with
matter messengers and positive squark masses squared)
could in principle have a minimum with matter messengers.
However, 
such a minimum 
does not connect smoothly
to the supersymmetric case when SUSY breaking vanishes. Therefore,
for small SUSY breaking, 
this is probably not the true vacuum of the dual theory.
For large SUSY breaking, this minimum could have lower
energy than the gauge messenger minimum we discussed 
so far, 
and a
phase transition in the magnetic theory could occur. 
There is, however, no good reason to expect that this vacuum 
gives the dual description
of the electric theory after the phase transition.

\section{One-loop contributions to the scalar masses}  %
\label{oneloop}

This appendix describes the calculation of
the one-loop contributions to the 
scalar masses. It also contains some explicit expressions
for these masses.

Let us first review in more detail the spectrum of the dual
theory  at tree level.
The gauge group $SU(\N+1)$ is higgsed down to $SU(\N)$
by the expectation values of $h$ and $\hb$.
Thus there are $2\N -1$ heavy gauge multiplets, corresponding
to the broken generators.
These  consist of 
vectors of mass $k g v$, scalars of mass 
$v\, \sqrt{ k g^2 + 2 \lambda^2 f^2 }$, 
and fermions of mass
$\pm v\,\left(\sqrt{
k g^2 +\lambda^2 f^2/4} \pm \lambda f/2 \right)$,
where $f \equiv F_h/v^2$, and $k$ is a group theory factor:
$k=1$ for $2(\N-1)$ of the vector multiplets,
and $k = 2(\N-1)/\N $ for one multiplet, corresponding
to the broken ``$U(1)$" generator.
Additional fields become heavy through their superpotential
coupling to $h$ and $\hb$. 
These include the mesons $V^i$, $\Vb_i$ and the ``broken" component
of the dual quarks and antiquark. 
Together, these combine to form $2\N$ chiral multiplets,
$V^i_{\pm}$, $\Vb_{i\mp}$,
 with  scalars of mass 
$m_\pm^2 = \lambda^2\, v^2\, (1\pm {f\over \lambda}) \ , $
and fermions of mass $\lambda v$.
Note that for $f > \lambda$, these fields obtain vevs, 
but we will not study this possibility here. 
We assume that some coupling of the DSB sector 
can be chosen small enough so that $f <\lambda$.

As mentioned above, the scalar
masses receive contributions starting at the one-loop level.
The one-loop contributions start at order
${\cal O}(f^4)$ for scalar masses squared, 
whereas the two-loop contributions
start at order ${\cal O}(f^2)$,
with $f\equiv F_h/v^2$.
 
Before going into  details, it is instructive to understand 
the vanishing of the leading-order one loop contributions~\cite{GR}. 
As the authors of~\cite{GR} point out, to leading order in the
 supersymmetry-breaking parameter,
the scalar masses are given by the {\it second} derivative of the relevant
wave-function renormalization with respect to the logarithm of the 
threshold scale. Since the one-loop contribution only involves
{\it single} logs, it vanishes to leading order in the supersymmetry breaking.

The scalar mesons then get masses at one-loop
through their superpotential couplings to 
$V^i_{\pm}$, $\Vb_{i\mp}$,
\beq
\label{mesononeloop}
\tilde{M}^2_{M} \ =\
{1\over 16\pi^2 }\,\lambda^4 \, v^2 \,
\left[ (2+{f\over\lambda})\,
\ln(2+{f\over\lambda})
\, + \,
(2 - {f\over\lambda})\,
\ln(2 - {f\over\lambda}) \right]
\ ,
\eeq
which is always negative.

This contribution dominates the two-loop 
contribution~(\ref{mesonmass})
for $f > \sqrt{\N} \ay$. 
Thus, there can be some intermediate region of $f$,
namely,   $\lambda> f > \sqrt{\N} \ay$,
for which this contribution is relevant.

The dual squark masses are more complicated. 
These receive contributions both through their superpotential 
couplings to the fields $V^i_{\pm}$, $\Vb_{i\pm}$,
and through their gauge and superpotential couplings to the
heavy vector multiplets.
The resulting masses can be written as,
\beq
\label{oneloopsq}
\tilde{M}^2_q \ =\
M^2_{U(1)} \ +\ {M^\prime}^2 \ ,
\eeq
where $M^2_{U(1)}$ is induced by the vector multiplet
corresponding to the broken ``$U(1)$" generator,
and is given by 
\beq
\label{muone}
M_{U(1)}^2\ =\ 
{1\over 16 \pi^2}\, {1\over \N^2}\, g_m^4\, v^2\,
\left[
\ln(1+8 f_n^2) \ -\
{8\, f_n\over
\sqrt{1+f_n^2} }\,
\ln( \sqrt{1+f_n^2\nonumber\\} + f_n) 
\right] \ ,
\eeq
where $f_n = \sqrt{\N/(2(\N-1))}\, f/g_m$. 

Interestingly,
$M_{U(1)}^2 < 0$ for all values of $f_n$.
Note that this contribution is a pure gauge contribution,
and is suppressed for large $\N$.

The remaining piece of the squark mass, 
${M^\prime}^2$, depends on both the gauge coupling and Yukawa coupling,
but is $\N$-independent.
It is given by a rather lengthy expression
\beqa
\label{mprime}
&{M^\prime}^2& \ =\ {1\over 16\pi^2}\lambda^4 \, v^2 
\, \times\, \Bigg[
\\~\nonumber
&\,&{\g}^4\,(\f\,(-\g^4+3\f^4+15\f^2\g^2-3\f^2\g^4-\g^6)\, 
\ln(\g)\, 
\\~\nonumber
 &\times& 
{
(\f+ \sqrt{4\g^2+\f^2}  )/2
-2\g^6+3f^4\g^2+12\f^2\g^4
-3\f^2\g^6+2\g^8
\over 
\f\,(5\f^2\g^2+5\g^4+\f^4-3\g^6-\f^2\g^4)\,
(\f+  \sqrt{4\g^2+\f^2}  )/2
+\g^2\,(4\f^2\g^2+2\g^4+\f^4-2\g^6-\f^2\g^4) }
\\~\nonumber
&+&\,
{(1+\f)\,(-2-\f+2\f^2+2\g^2-\f \g^2+\f^3+\f^2 \g^2)
\over 2\,
(-1-\f+\g^2+2 \f^2)}
\,\ln(1+\f)
\\~\nonumber
&+&\,
{(1-\f)\,(-2+\f+2\f^2+2\g^2+\f \g^2-\f^3+\f^2 \g^2)
\over 2\,
(-1+\f+\g^2+2 \f^2)}
\,\ln(1-\f)
\\~\nonumber
&+&
{(\g^2+2\f^2)\,
(\g^6-2\g^4+2\f^2 \g^4+
\g^2+3\f^2 \g^2-6\f^2+6\f^4)
\over 2
(-1-\f+\g^2+2\f^2)\,
(-1+\f+\g^2+2\f^2)
}\,\ln(\g^2+2\f^2) 
\\~\nonumber
&+& 4\ln(C) 
\\~\nonumber
&\times&
{\f^2\,(\f^2+2\g^2)\,
(-\g^8-\g^4+3\,\f^2 \g^2+2\g^6+\f^4)\,
C
-\f\,\g^2\,
(\g^2+\f^2)\,(\g^8+\g^4-3\f^2\,\g^2-2\g^6-\f^4)
\over
\,(-\f-1+\g^2)\,(\f-1+\g^2)\, (
(2\g^4+4\f^2 \g^2+\f^4)\, C
+\g^2 \f\,(\f^2+3\g^2) )
} \Bigg]
\\~\nonumber
\eeqa
where
$C = (\f+  \sqrt{4\g^2+\f^2} )/2$,
and $\f = f/\lambda$, $\g = g_m/\lambda$.



\begin{thebibliography}{99}

\bibitem{Seiberg}
N.~Seiberg, {\it Nucl. Phys.} {\bf B435}, 129 (1995);
for a review, see K.~Intriligator and N.~Seiberg,
{\it Nucl. Phys. Proc. Suppl.} {\bf 45BC}, 1 (1996).



\bibitem{Hsu}
N.~Evans, S.~D.~H.~Hsu, M.~Schwetz,
{\it Phys. Lett.} {\bf B355}, 475 (1995);
N.~Evans, S.~D.~H.~Hsu, M.~Schwetz, S.~B.~Selipsky,
{\it Nucl. Phys.} {\bf B456}, 205 (1995).

\bibitem{Hsu2}
N.~Evans, S.~D.~H.~Hsu, M.~Schwetz,
{\it Phys. Lett.} {\bf B404}, 77 (1997).

\bibitem{Peskin}
O.~Aharony, J.~Sonnenschein, M.E.~Peskin, S.~Yankielowicz,
{\it Phys. Rev.} {\bf D52}, 6157 (1995).

\bibitem{DHoker}
E.~D'Hoker, Y.~Mimura and N.~Sakai,
{\it Phys. Rev.} {\bf D54}, 7724 (1996).

\bibitem{terninganomaly}
Non-supersymmetric QCD may have  a dual description that can not
be obtained from the Seiberg dual.
Some possibilities, based on anomaly matching considerations are
listed in:
J.~Terning, 
{\it Phys. Rev. Lett.} {\bf 80}, 2517 (1998).

\bibitem{A-G}
L.~\'Alvarez-Gaum\'e, J.~Distler, C.~Kounnas and M.~Mari\~no,
{\it Int. J. Mod. Phys.} {\bf A11}, 4745 (1996);
L.~\'Alvarez-Gaum\'e and M.~Mari\~no, 
{\it Int. J. Mod. Phys.} {\bf A12}, 975 (1997);
L.~\'Alvarez-Gaum\'e, M.~Mari\~no and F.~Zamora,
{\it Int. J. Mod. Phys.} {\bf A13}, 403 (1998);
{\it Int. J. Mod. Phys.} {\bf A13}, 1847 (1998).

\bibitem{seibergwitten}
N.~Seiberg and E.~Witten,
{\it Nucl. Phys.} {\bf B426}, 19 (1994), Erratum {\it ibid.} {\bf B430}, 
485 (1994), and {\it Nucl. Phys.} {\bf B431}, 484 (1994).

\bibitem{nonchiralmap}
M.~Berkooz, 
{\it Nucl. Phys.} {\bf B466}, 75 (1996).

\bibitem{GMSB}
M.~Dine, W.~Fischler, and M.~Srednicki, Nucl.~Phys. {\bf B189}, 575
(1981);
C.~Nappi and B.~Ovrut, Phys.~Lett.~B {\bf 113}, 175 (1982);
M.~Dine and W.~Fischler, Nucl.~Phys. {\bf B204}, 346 (1982);
L.~Alvarez-Gaume, M.~Claudson and M.~Wise, Nucl.~Phys. {\bf B207}
96 (1982). 

\bibitem{dnns}
M.~Dine, A.~Nelson, and Y.~Shirman, Phys.~Rev.~D {\bf 51}, 
1362 (1995);
M.~Dine, A.~Nelson, Y.~Nir, and Y.~Shirman, Phys.~Rev.~D {\bf 53}, 
2658 (1996).

\bibitem{review}
For a review, see  G.~F.~Giudice and R.~Rattazzi, hep-ph/9801271.


\bibitem{PST}
E.~Poppitz, Y.~Shadmi and S.P.~Trivedi,
{\it Nucl. Phys.} {\bf B480}, 125 (1996).

\bibitem{nophasetransition}
K.~Intriligator and N.~Seiberg, 
{\it Nucl. Phys.} {\bf B431}, 551 (1994).

\bibitem{seibergexact}
N.~Seiberg,  
{\it Phys. Rev.} {\bf D49}, 6857 (1994).

\bibitem{DNS}
M.~Dine, Y.~Nir and Y.~Shirman, 
{\it Phys. Rev.} {\bf D55}, 1501 (1997).

\bibitem{GR}
G.~F.~Giudice and R.~Rattazzi,   
{\it Nucl. Phys.} {\bf B511}, 25 (1998).

\bibitem{BZ}
T.~Banks and A.~Zaks, {\it Nucl. Phys.} {\bf B196}, 189 (1982).

\bibitem{HS}
J.~Hisano and M.~Shifman, {\it Phys. Rev.} {\bf D56}, 5475 (1997).

\bibitem{nima}
N.~Arkani-Hamed, private communication.

\bibitem{CW}
S.~Coleman and E.~Weinberg, {\it Phys. Rev.} {\bf D7}, 1988 (1973). 

\bibitem{Terning}
T.~Appelquist, J.~Terning and L.~C.~R.~Wijewardhana,
{\it Phys. Rev. Lett.} {\bf 77}, 1214 (1996).

\bibitem{VW}
C.~Vafa and E.~Witten, {\it Nucl. Phys.} {\bf B234}, 173 (1984).




\end{thebibliography}
\end{document}